\documentclass[conference]{IEEEconf}
\usepackage{hyperref}
\usepackage{amsmath}
\usepackage{tabularx}
\usepackage{graphicx}
\usepackage{algorithm2e}

\usepackage{bm}
\usepackage{xcolor}
\usepackage{amsfonts}
\usepackage{amssymb}
\usepackage{verbatim}
\usepackage{upgreek}
\usepackage{graphicx}
\usepackage{epstopdf}
\usepackage{subfigure}
\usepackage{color}
\usepackage{amsmath}
\usepackage{algorithmic}
\usepackage{siunitx}
\usepackage{tikz}
\IEEEoverridecommandlockouts

\begin{document}
\title{\LARGE \bf A Human-optimized Model Predictive Control Scheme and Extremum Seeking Parameter Estimator for Slip Control of Electric Race Cars}

\author{Wytze de Vries, Jorn van Kampen, Mauro Salazar \thanks{Control Systems Technology section, Eindhoven University
of Technology, Eindhoven, The Netherlands (e-mail: \href{mailto:w.a.b.d.vries@student.tue.nl}{w.a.b.d.vries@student.tue.nl}; \href{mailto:j.h.e.v.kampen@tue.nl}{j.h.e.v.kampen@tue.nl}; \href{mailto: m.r.u.salazar@tue.nl}{ m.r.u.salazar@tue.nl};)}%
}

\maketitle

\begin{abstract}
	This paper presents a longitudinal slip control system for a rear-wheel-driven electric endurance race car. The control system integrates Model Predictive Control (MPC) with Extremum Seeking Control (ESC) to optimize the traction and regenerative braking performance of the powertrain. The MPC contains an analytical solution which results in a negligible computation time, whilst providing an optimal solution to a multi-objective optimization problem. The ESC algorithm allows continuous estimation of the optimal slip reference without assuming any prior knowledge of the tire dynamics. Finally, the control parameters are determined using a human-driven preference-based optimization algorithm in order to obtain the desired response. Simulation results and comparisons with other methods demonstrate the system’s capability to automatically determine and track the optimal slip values, showing stability and performance under varying conditions.
\end{abstract}

\section{Introduction}\label{Chapters:Introduction}
The motorsports industry has followed the global shift towards electrification of vehicles. Following the introduction of hybrid electric technologies in categories like Formula 1, Le Mans Hypercars, Le Mans Daytona Hybrids, and even the Dakar Rally, fully electric vehicles have now entered the highest levels of competition. These electric vehicles compete in series such as Formula E and Extreme E. A key distinction between racing with fully electric vehicles and their fossil-fueled counterparts is the significant energy limitation due to the low energy density of batteries, which restricts electric racing formats to short sprint races at lower speed.
Therefore, making optimal use of the energy available onboard is of paramount importance. \\
The vehicle is often limited by grip when being driven on the limit in order to minimize lap time. Optimizing the difference between the speed of the driven wheels and the road, also known as wheel slip, becomes essential in order to not waste any energy, both during acceleration and regenerative braking. Consequently, systems such as anti-lock braking systems and traction control become essential for improving vehicle performance. These control systems should require minimal computational effort due to electronic control unit (ECU) limitations and be easily tunable by race engineers. The knowledge of tire parameters a-priori are often used to enhance the performance of model-based control systems as they allow for a more accurate description of the dynamics. However, the tire parameters are often unavailable or inaccurate, as they change considerably over the course of a race. Against this backdrop, this paper presents a method which jointly optimizes the motor torque and estimates the tire parameters for optimal performance. The hyper-parameters for this control system are selected using a human-driven preference based optimization algorithm. 

\begin{figure}[] 
    \centering
    \includegraphics[trim={0cm 4cm 0cm 4cm},clip,width=\columnwidth]{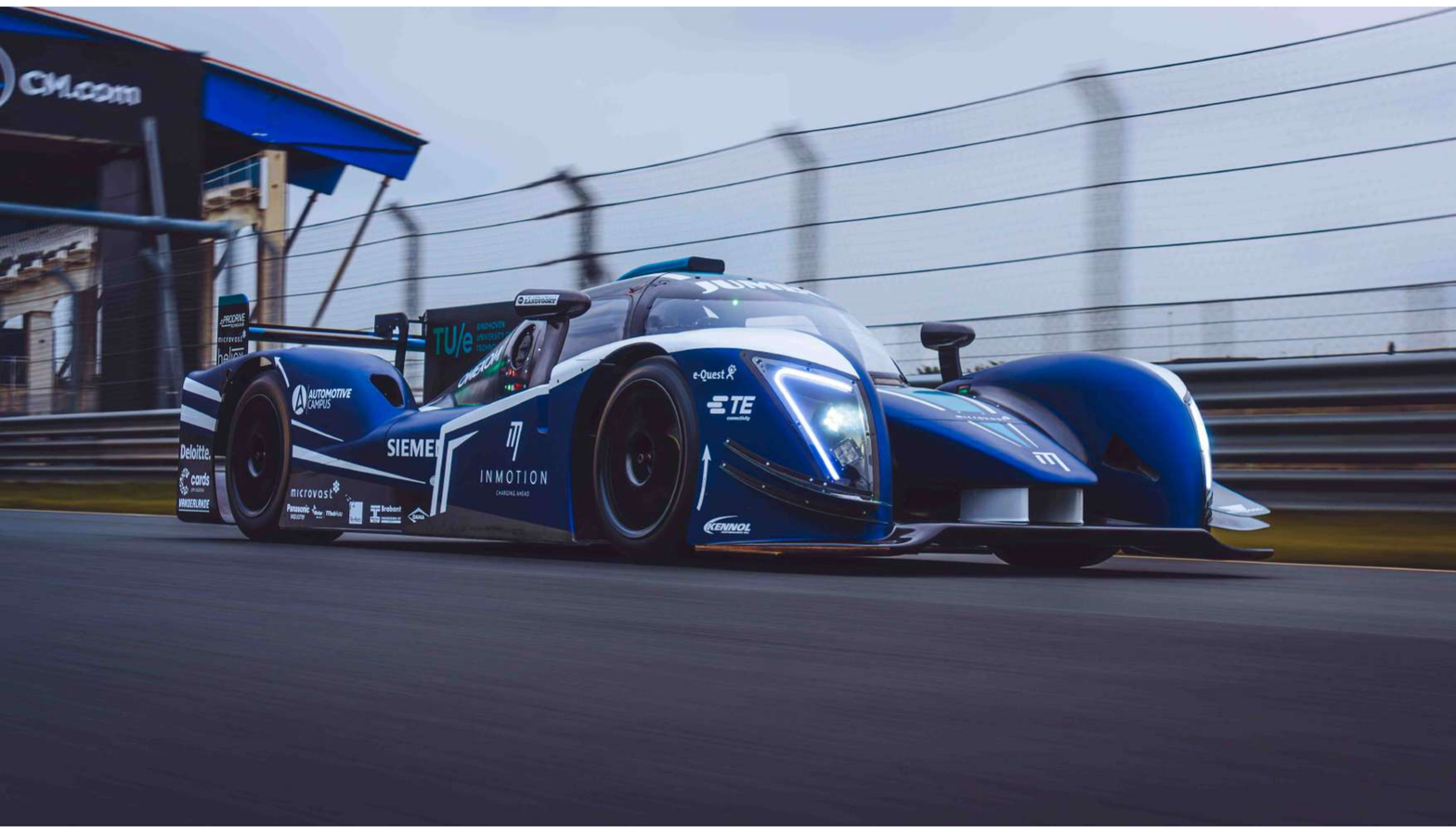}
    \caption{InMotion's fully electric endurance race car.}
    \label{fig:Introduction}
\end{figure}

\emph{Related literature}
This work pertains to three main research streams: Slip control methods which do not assume any knowledge of the tire parameters, those who do assume knowledge of the tire parameters, and slip reference optimization methods.
The first research stream includes approaches such as rule based control systems \cite{gerard2010adaptation} \cite{pasillas2006hybrid} \cite{ait2004class} and Fuzzy logic control \cite{mauer1995fuzzy}  \cite{layne1993fuzzy} \cite{khatun2003application} , which provide good performance and are robust to environmental changes. They are therefore the most common type within industry. However, their primary drawback is the large number of control parameters, which require careful tuning to obtain robust performance. Regular PID controllers \cite{li2009traction} \cite{li2012pid} \cite{will1998sliding} allow for continuous control and have shown to be capable of high performance. They do however require careful tuning and lack guarantees of optimality due to the lack of knowledge of the system dynamics. Sliding mode control has been shown to perform well \cite{wu2003simulated} \cite{buckholtz2002reference} and offers significant advantages such as robustness to parameter variation and disturbances. It does however suffer from high frequency oscillations which could excite unmodelled drivetrain dynamics. \\
The second research stream are the methods which do utilize information of the tires. Optimal control methods have been proposed such as LQR \cite{lqr}, which relies on local linearization and gain scheduling. Whilst being proven to be theoretically exponentially stable (using Lyapunov theory), it showed fundamental limitations on the achievable performance. Yoo and Wang \cite{yoo2007model} proposed a model predictive control method, which was validated using HIL experiments to perform well on a range of surfaces. It did however require a predefined longitudinal slip stiffness and significant computational resources to implement the controller in real time. To address the computation challenges, explicit MPC was proposed in \cite{tavernini2018explicit} \cite{tavernini2019explicit} to reduce the computational burden of solving an optimization problem online.  Although this approach showed good performance, it required a predefined tire model. Borrelli et al. \cite{borrelli2006mpc} introduced a MPC/Hybrid approach that approximated tire characteristics with a piecewise affine model. While effective, the performance was limited by the accuracy of the tire model approximation, and oscillations occurred due to the absence of a quadratic cost in the objective function. \\
The final research stream are methods which actively search for the optimal slip reference which is subsequently tracked by the slip controllers. Several authors \cite{will1998sliding} \cite{drakunov1995abs} proposed using sliding modes to estimate the slip reference which resulted in the largest acceleration. This was proven to be an effective method on regular vehicles. This method is however ill-suited to racing vehicles which use aerodynamics to generate more grip at high speeds. Thereby, downforce decreases drastically when the brakes are applied and the vehicle slows down. The maximum longitudinal acceleration will therefore decrease significantly during braking, which would mean that the sliding mode estimator would not be able to distinguish if the reduction in longitudinal acceleration is due to down-force or the slip reference provided. Nesic et al. \cite{nesic2012framework} proposed a method which leveraged extremum seeking control (ESC) to estimate the tire characteristics. The $\mu/\kappa$ curve was approximated using a predefined function which was parameterized by 5 parameters. ESC was then used to estimate these parameters online. Simulation studies showed that it was able to provide good estimations of the parameters if sufficient data was available. However, the achievable accuracy becomes bounded by using a predefined function to approximate the true $\mu/\kappa$ curve. Furthermore, the whole $\mu/\kappa$ curve is estimated while only the slip reference resulting in the peak force is required. \\
In conclusion, to the best of the authors knowledge, there is no optimal slip control method which does not require any prior knowledge of the tire, requires low computational effort and has limited control parameters.

\begin{figure}[] 
    \centering
    \includegraphics[trim={0cm 0cm 0cm -0.5cm},clip,width=\columnwidth]{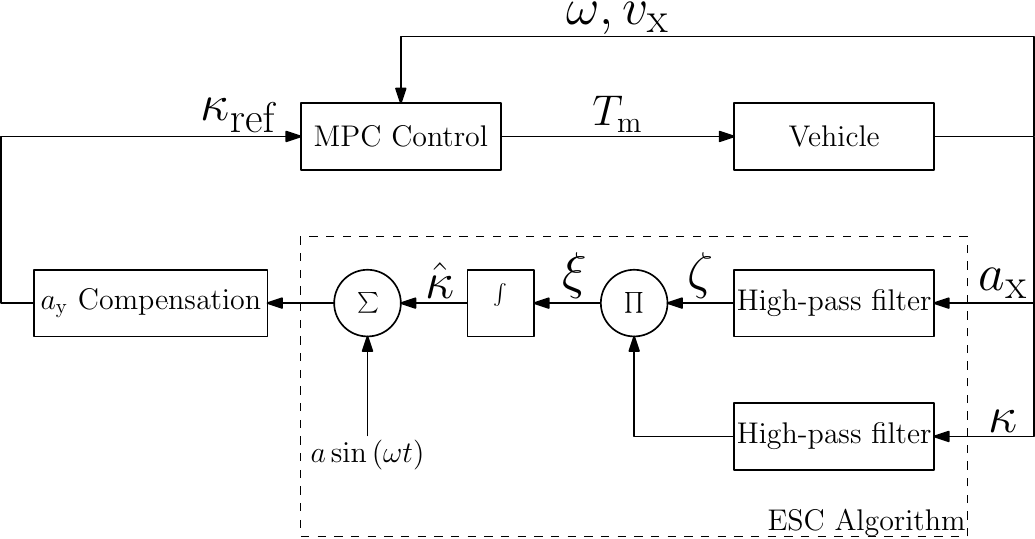}
    \caption{Simplified architecture of the implemented slip control strategy. The ESC Algorithm estimates the optimum slip through gradient estimation, which is subsequently scaled to account for lateral dynamics. The MPC uses information of the wheel speeds to control the slip of the rear tires by modulating the motor torque.}
    \label{fig:ESC}
\end{figure}

\emph{Statement of Contributions}
This paper introduces an unconstrained MPC with integral action to address the slip control problem. With an analytical solution, the MPC efficiently tracks slip references for each rear tire. Leveraging ESC to identify the optimal longitudinal slip associated with peak longitudinal force. ESC is used, as it actively perturbs the slip reference, and is therefore able to determine the optimal slip reference even in the presence of down-force. The slip reference is dynamically adjusted to maintain lateral stability. A schematic of the proposed method is shown in Fig.~\ref{fig:ESC}. Additionally, the control hyper-parameters are optimized using the human-driven preference learning algorithm c-GLISp \cite{cGLISp}. The control system is demonstrated using the electric endurance race car from student team InMotion of Eindhoven University of Technology shown in Fig.~\ref{fig:Introduction}. The vehicle is rear-wheel driven with a single electric motor, coupled with a single-speed gearbox and a limited slip differential. 

\emph{Organization}
The paper is organized as follows. The MPC scheme and ESC algorithm will be presented in Section.~\ref{Chapters:MPC} and Section.~\ref{Chapters:ESC} respectively. The optimization of the control parameters will be discussed in Section.~\ref{Chapters:cGLISp} and the resulting performance will be shown in Section.~\ref{Chapters:Results}. Finally the paper will be concluded in Section.~\ref{Chapters:Conclusion}.

\section{MPC}\label{Chapters:MPC}
In this section we will first review the relevant wheel slip dynamics, after which we will continue with leveraging these dynamics into a MPC problem.
\subsection{Modeling}
\begin{figure}[] 
    \centering
    \includegraphics[trim={0cm 0cm 0cm -0.5cm},clip,width=0.5\columnwidth]{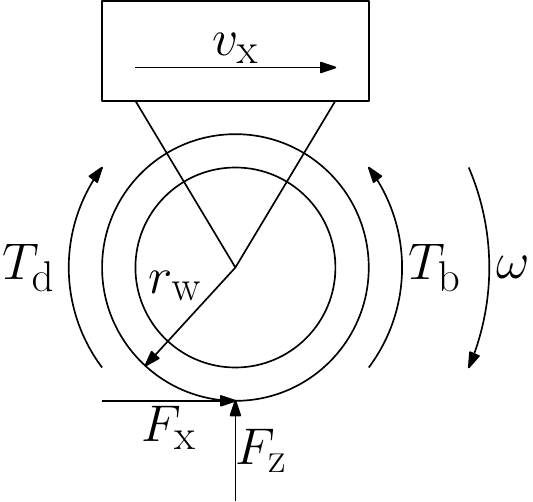}
    \caption{Quarter vehicle model of a vehicle traveling at the speed of $v_{\mathrm{x}}$, which exerts a vertical force $F_\mathrm{z}$ on the wheel. The wheel has a radius $r_\mathrm{w}$, and speed $\omega$, which is dependent on the brake torque $T_\mathrm{b}$, drive torque $T_\mathrm{d}$ and longitudinal force $F_\mathrm{x}$ }
    \label{fig:wheel_dyn}
\end{figure}

Given a car with wheel speed $\omega$, effective wheel radius $r_{\mathrm{w}}$, and inertia $I_\mathrm{xx}$, the resulting wheel dynamics are described by
\begin{equation}
    I_\mathrm{xx} \cdot \Dot{\omega} = T_\mathrm{d} - T_\mathrm{b} - r_{\mathrm{w}}\cdot F_{\mathrm{x}}(\kappa, \alpha, \mu, F_{\mathrm{z}}),
\end{equation}
where $T_\mathrm{d}$ is the drive torque applied to the wheel, $T_\mathrm{b}$ is the brake torque applied to the wheel, and $F_{\mathrm{x}}$ is the longitudinal force produced by the tire. This longitudinal force depends on a number of parameters, such as the longitudinal slip ratio $\kappa$, the slip angle $\alpha$, the road friction coefficient $\mu$, and the vertical load exerted on the tire $F_{\mathrm{z}}$. A visual representation is shown in Fig.~\ref{fig:wheel_dyn}.
During braking, the longitudinal slip ratio is defined by
\begin{equation}
    \kappa = \frac{\omega \cdot r_{\mathrm{w}} - v_{\mathrm{x}}}{v_{\mathrm{x}}} = \frac{v_{\mathrm{slip}}}{v_{\mathrm{x}}},
    \label{eq:kappa1}
\end{equation}
where $v_{\mathrm{x}}$ is the vehicle speed and $v_{\mathrm{slip}}$ is the difference in speed between the road and tire at the tire contact patch. And under acceleration the longitudinal slip ratio is defined by 
\begin{equation}
    \kappa = \frac{\omega \cdot r_{\mathrm{w}} - v_{\mathrm{x}}}{\omega \cdot r_{\mathrm{w}}} = \frac{v_{\mathrm{slip}}}{\omega \cdot r_{\mathrm{w}}},
    \label{eq:kappa2}
\end{equation}
which is often expressed as a percentage.
The torque applied to the wheel is dependent on the input torque provided by the motor and the dynamics of the limited slip differential. The torque applied to each wheel is then described by
\begin{equation}
\begin{split}
    &T_\mathrm{d,L} = \frac{\gamma}{2} \cdot T_\mathrm{m} + T_\mathrm{c}(\omega_L, \omega_R, T_\mathrm{m}), \\
    &T_\mathrm{d,R} = \frac{\gamma}{2} \cdot T_\mathrm{m} - T_\mathrm{c}(\omega_L, \omega_R, T_\mathrm{m}),
\end{split}
\end{equation}
where $T_\mathrm{d,L}$ and $T_\mathrm{d,R}$ are the drive shafts torques for the left and right wheel respectively, $\gamma$ is the gear ratio, $T_\mathrm{m}$ is the motor torque and $T_\mathrm{c}$ is the torque developed by the clutch of the limited slip differential. The clutch torque is dependent on the difference between the wheel speeds and the torque applied to the differential. The MPC problem can be constructed now that all the relevant dynamics have been introduced. 

\subsection{MPC scheme}
In this section we will discuss how we define the prediction model and the objective function for the MPC problem. The dynamics presented in the previous section can be discretized using a forward Euler method and then combined into the state space form of
\begin{equation}
\begin{split}
    \begin{bmatrix}
        \omega_\mathrm{L}(k+1) \\ 
        \omega_\mathrm{R}(k+1) \\
        v_\mathrm{x}(k+1) \\
    \end{bmatrix}
     = & 
    \underbrace{I_\mathrm{3}}_{A_\mathrm{p}}
    \begin{bmatrix}
        \omega_\mathrm{L}(k) \\ 
        \omega_\mathrm{R}(k) \\
        v_\mathrm{x}(k) \\
    \end{bmatrix}
    + 
    \underbrace{\begin{bmatrix}
        \frac{T_\mathrm{s} \cdot \gamma}{2 \cdot I_\mathrm{xx}} \\
        \frac{T_\mathrm{s} \cdot \gamma}{2 \cdot I_\mathrm{xx}} \\
        0 \\
    \end{bmatrix}}_{B_\mathrm{p}}
    T_\mathrm{m} \\ 
    & + 
    \underbrace{\begin{bmatrix}
        -\frac{T_\mathrm{s} \cdot r_\mathrm{w}}{I_\mathrm{xx}} & 0  & \frac{T_\mathrm{s}}{I_\mathrm{xx}}\\
        0 & -\frac{T_\mathrm{s} \cdot r_\mathrm{w}}{I_\mathrm{xx}} & -\frac{T_\mathrm{s}}{I_\mathrm{xx}}\\
        \frac{T_\mathrm{s}}{m} & \frac{T_\mathrm{s}}{m} & 0 \\
    \end{bmatrix}}_{B_\mathrm{d}}
    \begin{bmatrix}
        F_\mathrm{{x_L}} \\
        F_\mathrm{{x_R}} \\
        T_\mathrm{c}  \\
    \end{bmatrix},
    \\
    \begin{bmatrix}
        v_\mathrm{{slip,L}}(k) \\
        v_\mathrm{{slip,R}}(k) \\
    \end{bmatrix}
    & =
    \underbrace{\begin{bmatrix}
        r_\mathrm{w} & 0 & -1 \\
        0 & r_\mathrm{w} & -1 \\
    \end{bmatrix}}_{C_\mathrm{p}}
    \begin{bmatrix}
        \omega_\mathrm{L}(k) \\ 
        \omega_\mathrm{R}(k) \\
        v_\mathrm{x}(k) \\
    \end{bmatrix},
\end{split}
\label{eq:MPC_int_State_space}
\end{equation}
where $T_\mathrm{s}$ is the sample time, and where $v_\mathrm{{slip,L}}$ and $v_\mathrm{{slip,R}}$ are the outputs. The slip velocities were used instead of slip ratio's in order to preserve linearity in the equations.  The state space model from~\eqref{eq:MPC_int_State_space} can be written in compact form:
\begin{equation}
\begin{split}
    x_{\mathrm{p}}(k+1) &:= A_{\mathrm{p}} x_{\mathrm{p}}(k) + B_{\mathrm{p}} u(k) + B_{\mathrm{d}} d(k), \\
    y(k) &:= C_{\mathrm{p}} x_{\mathrm{p}}(k),
\end{split}
\label{eq:state_p}
\end{equation}
where $x_{\mathrm{p}}$ is the state vector, $u$ is the input, $d$ are the unmeasured disturbances, and $y$ are the outputs. The longitudinal force generated by the tires is treated as a constant disturbance, as the MPC formulation includes integral action capable of compensating for such constant disturbances. Any changes in the longitudinal force due to longitudinal slip, slip angle, suspension dynamics, load transfer or downforce will be considered unmeasured disturbances. Similarly, the clutch torque will also be regarded as an unmeasured disturbance. These disturbances will be neglected in the design of the MPC. \\
Our objective is to track a reference output (the slip velocities) without input oscillations which could upset the vehicle dynamics or damage the drivetrain. Therefore we use the standard truncated LQR formulation:
\begin{equation}
\begin{split}
     J(x(k),\Delta U_\mathrm{k}) =~&(y_\mathrm{{N|k}} - r_\mathrm{{N|k}})^\top P (y_\mathrm{{N|k}} - r_\mathrm{{N|k}}) \\ 
     &+ \sum_{i=0}^{N-1}( (y_\mathrm{{i|k}} - r_\mathrm{{i|k}})^\top Q (y_\mathrm{{i|k}} - r_\mathrm{{i|k}}) \\
     &+ \Delta u_\mathrm{{i|k}}^\top R \Delta u_\mathrm{{i|k}}),
\end{split}
\label{eq:cost1}
\end{equation}
which uses positive semi definite matrices $P$, $Q$, and $R$. The prediction horizon is denoted by $N$, and $r$ are the reference slip velocities. The output reference $r$ can be computed using (\ref{eq:kappa1}),(\ref{eq:kappa2}) and the reference slip ratio $\kappa_\mathrm{ref}$, which will be introduced in the next section. An analytical solution can be found due to the unconstrained nature of the problem. The derivation of this solution is discussed in the appendix. This means that we have obtained an optimal solution for output reference tracking, which can be implemented through direct state feedback at a negligible computational cost.  The only parameters which require tuning are the prediction horizon $N$, and the matrices $P$ and $Q$, while $R$ will remain fixed at 1. How we select these hyper-parameters will be discussed in Section.~\ref{Chapters:cGLISp}. However, we will first discuss how we determine the optimum slip reference $\kappa_\mathrm{ref}$ to maximize performance in the next Section.

\section{Extremum Seeking Control}\label{Chapters:ESC}
\begin{figure}[] 
    \centering
    \includegraphics[trim={3.8cm 11cm 4.2cm 11.6cm},clip,width=\columnwidth]{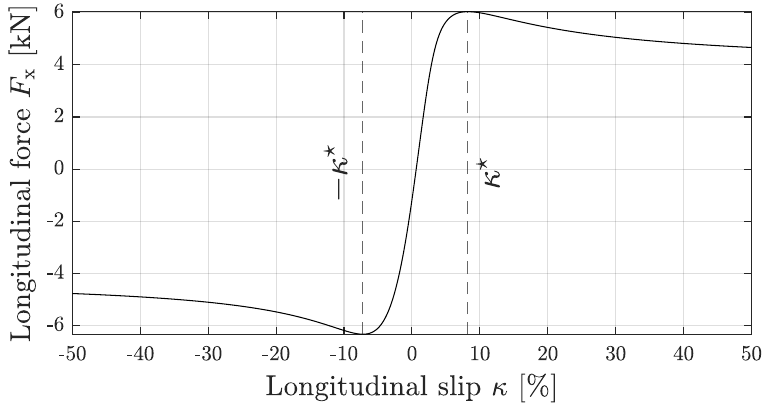}
    \caption{An example of the relation between the longitudinal slip and the longitudinal force produced by the tire.}
    \label{fig:FxKappaCurve}
\end{figure}

Determining the longitudinal slip reference $\kappa_\mathrm{ref}$ that maximizes the longitudinal force is a time-varying, non-linear yet convex optimization problem. The optimization problem corresponds to finding the location of the maximum of the $\kappa/F_\mathrm{x}$ curve as shown in Fig.~\ref{fig:FxKappaCurve}. The exact shape of this curve can vary based on factors such as the type of tire used, the tire slip angle, the temperature and pressure of the tire, the asphalt, the track conditions, and the load exerted on the tire. The curve is assumed to be unknown due to the large uncertainty.
ESC was therefore selected to solve this optimization problem. ESC is an adaptive, real-time optimization technique used to drive a system to an optimum point where the performance is maximized. In this case, ESC is used to find the longitudinal slip reference, which maximizes the longitudinal force and therefore longitudinal acceleration. This is done by perturbing the slip reference, which is tracked in real-time by the MPC, and subsequently directly measuring the longitudinal acceleration. Using this information we can estimate the gradient of the acceleration with respect to the longitudinal slip. We can use this estimated gradient to determine in which direction to seek the slip reference which maximizes the longitudinal acceleration. One key advantage of using ESC is its ability to directly search for the reference slip value that yields the maximum acceleration. This process is unaffected by estimation bias or steady-state tracking error as these inaccuracies do not impact of the quality of the optimal solution. 

ESC can be applied to our problem as follows. We first assume that the tire is symmetrical for positive and negative slip values. Therefore 
\begin{equation}
    F_\mathrm{x}(\kappa) = -F_\mathrm{x}(-\kappa),
\end{equation}
where $F_\mathrm{x}$ is the longitudinal force produced by the tire. Therefore we will only work with positive values for slip when discussing the ESC algorithm. The sign of the slip reference provided to the MPC is reversed during braking. We define the optimal slip $\kappa^\star$ as
\begin{equation}
    F_\mathrm{x}(\kappa^\star) \geq  F_\mathrm{x}(\kappa)~\forall\kappa.
\end{equation}
This optimal slip can be obtained by continuously estimating and subsequently integrating the gradient $\partial_\kappa F_\mathrm{x}(\kappa)$. 
However, the accuracy of the gradient estimation can be enhanced, as proposed in \cite{skafte2017introduction}. The original method assumes no phase shift between the input (slip reference $\kappa_\mathrm{ref}$) and the performance criterion (longitudinal acceleration $a_\mathrm{x}$). The proof of this statement requires a short review on how the gradient is estimated.
The current estimate of the optimal slip $\hat{\kappa}$ is perturbed by a sinusoidal signal resulting in the reference signal $\kappa_\mathrm{ref}$
\begin{equation}
    \kappa_\mathrm{ref} = \hat{\kappa} + a \cdot \sin{(\omega_\mathrm{p} t)},
\end{equation}
where $a$ is the amplitude of the perturbation, $\omega_\mathrm{p}$ is the frequency of the perturbation, and $t$ is time. This reference slip is provided to the MPC, which will track this time-varying reference. The resulting change in slip of the tires will result in a change in acceleration of the vehicle. This relation between the reference slip and the resulting longitudinal acceleration is represented by $g$. The resulting longitudinal acceleration $a_\mathrm{x}$ is then measured using an acceleration sensor. Using a first-degree Taylor expansion, this can be approximated as
\begin{equation}
\begin{split}
    a_\mathrm{x} &= g(\kappa_\mathrm{ref})\\ &= g(\hat{\kappa}^\star) + \partial_{\kappa_\mathrm{ref}} g(\kappa_\mathrm{ref}) \cdot a \cdot \sin{(\omega_\mathrm{p} t + \theta)},
\end{split}
\end{equation}
where the second term approximates the gradient of $g$, and $\theta$ is the phase lag introduced by the controller, dynamics and measurement and transmission delays. A second order high pass filter $H$, with a bandwith of $\omega_\mathrm{p}$, is applied to remove the DC component yielding the intermediate signal
\begin{equation}
    \zeta = H(a_\mathrm{x}) = \partial_{\kappa_\mathrm{ref}} g(\kappa_\mathrm{ref}) \cdot A \cdot a \cdot \sin{(\omega_\mathrm{p} t + \theta + \varphi)},
\end{equation}
where $A$ and $\varphi$ are the gain and phase shift introduced by the high pass filter. The gradient estimate $\xi$ can then be reconstructed by multiplying $\zeta$ with the original perturbation signal. However, a cleaner gradient estimate can be obtained when multiplying $\zeta$ by the high pass filtered estimated slip $\Tilde{\kappa}$. The estimated slip is obtained using the wheel speeds measured on the vehicle which are measured at the same time and therefore contain the same phase lag $\theta$. The resulting gradient estimate $\xi$ is then defined by:
\begin{equation}
\begin{split}
    \xi &= H(\Tilde{\kappa})\\
    &= H(\hat{\kappa} + a \sin{(\omega_\mathrm{p} t + \theta)}) \cdot \zeta \\ &= a^2 \cdot A^2 \cdot \sin^2{(\omega_\mathrm{p} t + \theta + \varphi)} \cdot \partial_{\kappa_\mathrm{ref}} \cdot g(\kappa_\mathrm{ref})\\ &= P(t) \cdot \partial_{\kappa_\mathrm{ref}} g(\kappa_\mathrm{ref}) ,
    \end{split}
\end{equation}
where $P(t)$ is always non-negative, offering a more reliable and cleaner gradient estimation compared to when the original perturbation signal is used which does not contain the phase lag $\theta$. \\
Convergence can be greatly improved by restricting the feasible search space to a predefined range of longitudinal slip references which can be illustrated using Fig.~\ref{fig:FxKappaCurve}. For instance, the slip was constrained to only include positive values, making the problem strictly convex. Excessively high slip values with small the gradients were excluded as well. A saturating integrator is used to enforce this restriction. \\
Lateral dynamics must be considered particularly for rear-wheel-drive vehicles, as the vehicle can become unstable if the rear tires experience excessive slip. The longitudinal slip reference should therefore be reduced whenever the vehicle experiences significant lateral acceleration. This relation should strike a compromise between stability and rotation in the corners, while maximizing tractive performance. It is difficult to define analytically due to the very sensitive dynamics related to vehicle balance during limit handling scenarios. The relationship can be refined in the future through experimentation, incorporating the subjective preferences of the driver. In the present paper, we opt for a simple approximation by using a piecewise affine map for scaling of the slip reference with the lateral acceleration $a_\mathrm{y}$, which can be expressed as follows:
\begin{equation}
\begin{cases}
    |a_\mathrm{y}|>a_\mathrm{0} &\rightarrow \kappa_{\mathrm{ref}} = 0 \\
    |a_\mathrm{y}|<a_\mathrm{\hat{\kappa}} &\rightarrow \kappa_{\mathrm{ref}} = \hat{\kappa} \\
    \textnormal{else} &\rightarrow \kappa_{\mathrm{ref}} = \frac{a_\mathrm{0}-|a_\mathrm{y}|}{a_\mathrm{0}-a_\mathrm{\hat{\kappa}}} \hat{\kappa}, \\
\end{cases}
\end{equation}
where $a_\mathrm{0}$ and $a_\mathrm{\hat{\kappa}}$ are parameters which can be selected in real time by the driver. The parameter $a_\mathrm{\hat{\kappa}}$ represents the lateral acceleration from which on the slip reference should start to be reduced, and the parameter $a_\mathrm{0}$ represents the lateral acceleration where the slip reference should have been reduced to zero. These parameters can be increased to improve the rotation of the car under braking and acceleration or decreased in order to improve stability. An overview of the discussed methods are connected is shown in Fig.~\ref{fig:ESC}. The hyper-parameters should be tuned before the control scheme is implemented online. Such an effort can be very expensive and is guided by the qualitative assessment of a human engineer. In order to speed op this process we use a human-based preference learning method which we will presented in the next section.

\section{Hyper-Parameter Optimization}\label{Chapters:cGLISp}
Now that the controllers have been introduced, the next step is to determine the hyper-parameters $P$, $Q$ and $N$. No clear objective is available as it instead involves multiple conflicting objectives, such as tracking error, overshoot, braking distance, disturbance rejection, damping, and others. It can be extremely challenging to capture all these factors in a single analytical objective function.\\
Therefore, we use a human-driven preference-based optimization method called c-GLISp \cite{cGLISp}. C-GLISp optimizes parameters by learning from comparisons evaluated by humans rather than requiring exact numerical objective function evaluations. The user compares two parameter sets and determines which one is preferable. The algorithm uses this preference feedback to build a surrogate model of the objective function. It is also able to handle unknown constraints as the user will provide feedback on the stability of the solution as well. By using surrogate models for both the objective function and the constraints, C-GLISp can iteratively propose new parameter sets, balancing exploration of new area's of the feasible space with the improvement of promising ones. The algorithm iterates through this process, continuously updating its models until it converges to an optimal solution. \\
This method is used to determine the parameters for the MPC which was introduced Section.~\ref{Chapters:MPC}. The parameters to be tuned are $P$, $Q$ and $N$, while $T_\mathrm{s}$ is minimized to 5 \unit{ms} as it is constrained by ECU limitations.

\section{Results}\label{Chapters:Results}
\begin{figure}[] 
    \centering
    \includegraphics[trim={2.5cm 9.5cm 2.5cm 9.6cm},clip,width=\columnwidth]{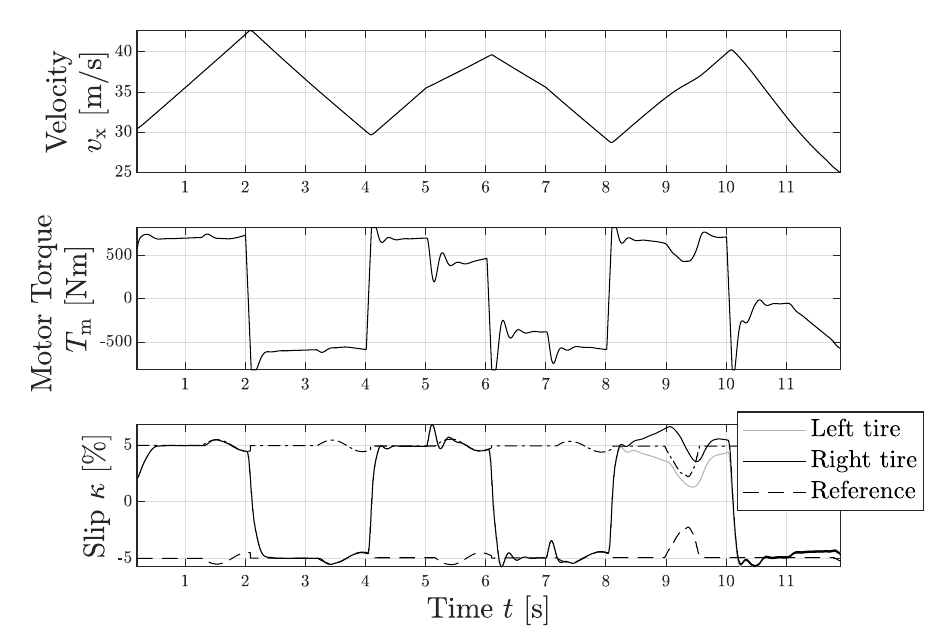}
    \caption{Simulation maneuver which was used to compare the parameter sets for c-GLISp with respect to tracking error, disturbance rejection, overshoot and damping. The simulation included a range of disturbances, such as changes in friction coefficient (at 5 seconds and 7 seconds), steering input (between 8 and 10 seconds) and braking input (between 10 and 12 seconds).}
    \label{fig:cGLISpTuning}
\end{figure}

This section presents simulation results from the implemented controller scheme, which was tuned using the c-GLISp method, and compares them with the approach proposed in \cite{will1998sliding}. This approach was selected for comparison as it employs a similar control architecture as proposed in this paper. Their method used a PID controller for reference tracking, combined with sliding modes to estimate the optimal slip reference.\\
The simulation environment included a 3 degree of freedom vehicle, load transfer, four individual wheels which used the Magic Tire formula version 5.2 models, road loads, downforce, a limited slip differential, drivetrain dynamics, and included measurement and actuation delays. The estimated slip that was used as an input for the controller was estimated through wheel speeds and a fixed tire radius was assumed.\\
The MPC control system was activated whenever the slip of one of the rear tires exceeded the reference slip, and deactivated whenever the driver requested less torque than the output of the control system. The ESC algorithm was activated after the MPC control system was active for more than 1\unit{s}, and deactivated whenever the lateral acceleration exceeded 1\unit{\frac{m}{s^2}}, whenever the brakes were applied or the MPC control was turned off. The perturbation signal of the ESC algorithm used a frequency of 1\unit{Hz} and an amplitude of 0.5\unit{\%}. The chosen frequency is significantly higher than the time-varying dynamics of the plant, yet remains within the bandwidth of the MPC. The amplitude selection represents a balance between achieving reliable convergence and minimizing oscillations.\\
\emph{Hyper-parameter optimization} Both parameter sets which were compared were simulated for each iteration of the c-GLISp method. Each simulation consisted of a sequence of repeated accelerating and braking events with multiple disturbances present, including changes in friction coefficient, steering input and braking input. A simulation example is shown in Fig.~\ref{fig:cGLISpTuning}. We evaluated the results based on tracking error, overshoot, disturbance rejection and damping. After 50 iterations, the resulting values were 250, 250 and 1450, for $P$, $Q$ and $N$ respectively. This approach allowed the user to focus on the desired behavior without requiring detailed knowledge of how each parameter influences the behavior of the controller. 

\begin{figure}[] 
    \centering
    \includegraphics[trim={3.2cm 9.55cm 4.3cm 9.6cm},clip,width=\columnwidth]{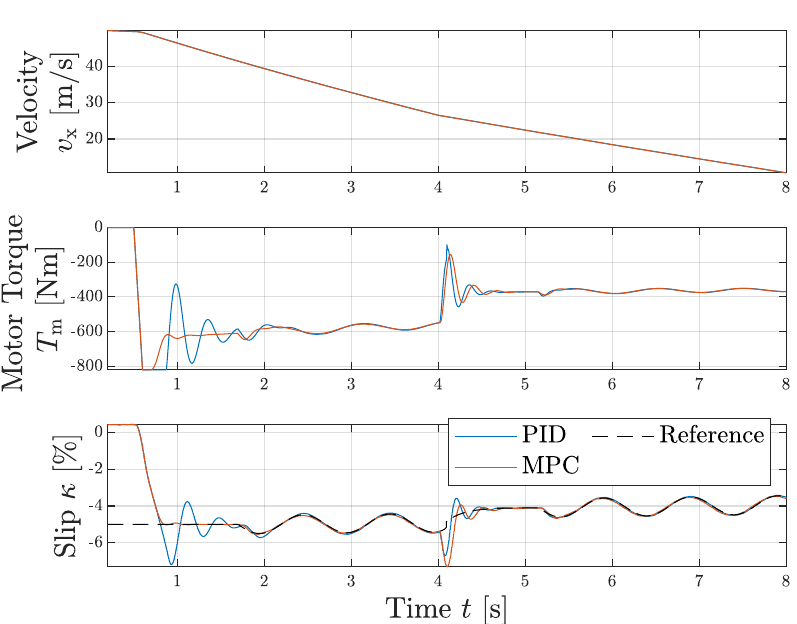}    \caption{Straight line braking maneuver with $\mu=0.6$ for $t\leq4$\unit{s} and $\mu=0.4$ for $t>4$\unit{s}. The MPC is shown to be able to track the slip reference with limited overshoot and oscillations, while the PID controller shows significant overshoot and oscillations.}
    \label{fig:MPC_results}
\end{figure}

\emph{MPC} The performance of the MPC is illustrated through a straight-line braking maneuver with a reduction in the road friction coefficient at 4\unit{s}. Figure~\ref{fig:MPC_results} shows the results for both the MPC and PID controllers, demonstrating that the MPC exhibits no overshoot and negligible oscillations in its initial response. In contrast, the PID controller overshoots the reference by 2.2\unit{\%} and displays significant oscillations. However, the PID adapts quicker to the friction change, with a 1.5\unit{\%} overshoot compared to 2.2\unit{\%} for the MPC.

\begin{figure}[] 
    \centering
    \includegraphics[trim={3.6cm 9.45cm 4.3cm 9.6cm},clip,width=\columnwidth]{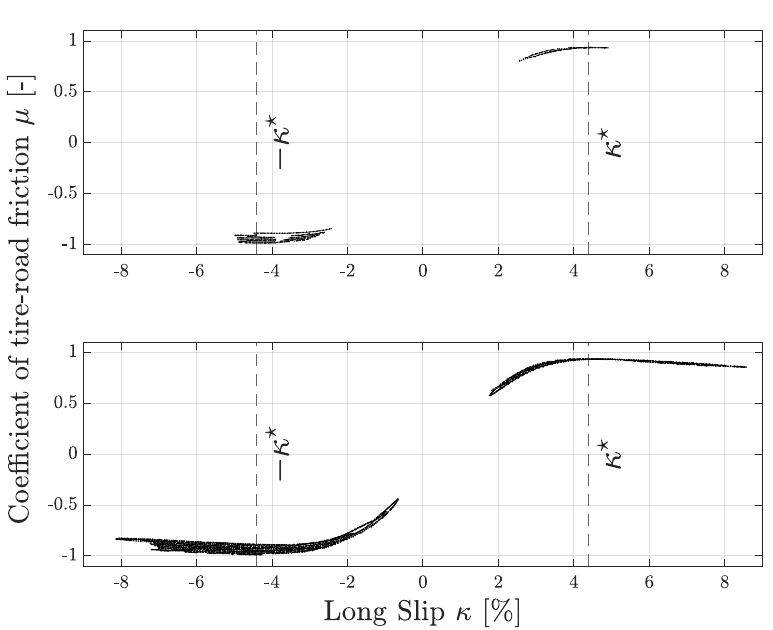}
    \caption{Slip control operating points of both tires on the $\kappa$/$\mu$ curve, for a sequence of straight line braking and acceleration events on a surface with a friction coefficient $\mu$ of 0.6.The ESC (top) is able to quickly converge to the optimal operating point $\kappa^\star$, while the sliding mode (bottom) shows a wider distribution around the optimal operating point due to significant oscillations.}
    \label{fig:ESC_results}
\end{figure}

\emph{ESC} The performance of the Extremum Seeking Control (ESC) algorithm is illustrated in Fig.~\ref{fig:ESC_results}. The simulation involved repeated acceleration and braking events between 20\unit{m/s} and 60\unit{m/s} over a duration of 100\unit{s}. Starting from an initial optimal slip estimate of 3\unit{\%}, the ESC converged to within 0.25\unit{\%} of the correct value (4.4\unit{\%}) after just two braking and two acceleration cycles. The slip controller consistently operated near the optimal points. In contrast, the sliding mode estimator required aggressive tuning to maintain stability and achieve convergence in the presence of strong disturbances, such as downforce, leading to significant and undesirable oscillations in the reference signals and system dynamics. These results indicate that the active perturbation used in the ESC algorithm enhances robustness against disturbances, stabilizing the estimation process.

\section{Conclusion}\label{Chapters:Conclusion}
This paper proposed a longitudinal slip control system for a rear-wheel-drive, electric endurance race car. The control system enabled the vehicle to maximize its efficiency by minimizing energy lost due to wheel spin. This was achieved by controlling the wheel slip using Model Predictive Control (MPC) and jointly optimizing the tire parameters by leveraging Extremum Seeking Control (ESC). The MPC problem contained an analytical solution which resulted in a negligible computation time while requiring few parameters to be tuned.  Additionally, the preference-based optimization of the hyper-parameters using the c-GLISp algorithm allowed for effective tuning without requiring extensive experience. Our results showed that the c-GLISp yielded hyper-parameters which resulted in desirable behavior. Moreover, the results showed that the MPC is able to accurately track the references without oscillation, while the ESC algorithm ensured that the performance of the tire is maximized.\\
Future work should include the implementation of the control scheme into the vehicle and performing real-life testing to validate the simulation study. Further investigation into the combined dynamics which are present during cornering should yield significant improvements in vehicle performance. 

\bibliographystyle{IEEEtran}
\bibliography{Bibliography}

\appendix
The objective function \ref{eq:cost1} requires the prediction model to provide the outputs as functions of the input rate. This is achieved by rewriting \ref{eq:state_p} into
\begin{equation}
\begin{split}
    \begin{bmatrix}
        \Delta x_{\mathrm{p}}(k+1) \\
        y(k+1) \\
    \end{bmatrix}
    &= 
    \begin{bmatrix}
        A_{\mathrm{p}} & 0 \\
        C_{\mathrm{p}} A_{\mathrm{p}} & I_\mathrm{2} \\
    \end{bmatrix}
        \begin{bmatrix}
        \Delta x_{\mathrm{p}}(k) \\
        y(k) \\
    \end{bmatrix}
    +
    \begin{bmatrix}
        B_{\mathrm{p}} \\
        C_{\mathrm{p}} B_{\mathrm{p}} \\
    \end{bmatrix}
    \Delta u(k), \\
    y(k) &= 
    \begin{bmatrix}
        0_{\mathrm{2x3}} & I_\mathrm{2} \\
    \end{bmatrix}
   \begin{bmatrix}
        \Delta x_{\mathrm{p}}(k) \\
        y(k) \\
    \end{bmatrix}
    \label{eq:MPC_delta_u}
\end{split}
\end{equation}
with $I$ denoting an identity matrix. And with $\Delta x_{\mathrm{p}}$ defined as
\begin{equation}
    \Delta x_{\mathrm{p}}(k) := x_{\mathrm{p}}(k)-x_{\mathrm{p}}(k-1),
\end{equation}
and $\Delta u$ defined as
\begin{equation}
    \Delta u(k) := u(k)-u(k-1).
\end{equation}
After which we can write~\eqref{eq:MPC_delta_u} into compact form
\begin{equation}
\begin{split}
    x(k+1) &= Ax(k) + B \Delta u(k), \\
    y(k) &= C x(k).
\end{split}
\end{equation}
We can predict the future outputs using the following prediction model
\begin{equation}
\begin{split}
    &\begin{bmatrix}
        y_{1|k}\\
        y_{2|k} \\
        \vdots \\
        y_{N|k} \\
    \end{bmatrix}
    = 
    \begin{bmatrix}
        CA \\
        CA^2 \\
        \vdots \\
        CA^N \\
    \end{bmatrix}
    x_{0|k} \\&+ 
    \begin{bmatrix}
        CB & 0 & \hdots & 0 \\
        CAB & CB & \hdots & 0 \\
        \vdots & \vdots & \ddots & \vdots \\
        CA^{N-1}B & CA^{N-2}B & \hdots & CB \\
    \end{bmatrix}
    \begin{bmatrix}
        \Delta u_{0|k} \\
        \Delta u_{1|k} \\
        \vdots \\
        \Delta u_{N-1|k} \\
    \end{bmatrix},        
\end{split}
\end{equation}
where $x_{0|k}$ is the current state. This can be written into compact form by
\begin{equation}
    Y_\mathrm{k} = \Phi x(k) + \Gamma \Delta U_\mathrm{k}
    \label{eq:pred_model}
\end{equation}
This allows us to rewrite~\eqref{eq:cost1} into
\begin{equation}
\begin{split}
    J(x(k),\Delta U_\mathrm{k}) = &x(k)^\top C^\top Q C x(k)\\ &+ (Y_\mathrm{k} - \Re_\mathrm{k})^\top \Omega (Y_\mathrm{k} - \Re_\mathrm{k})+ \Delta U_\mathrm{k}^\top \Psi \Delta U_\mathrm{k},
\end{split}
\end{equation}
with $\Re_\mathrm{k}$ being a matrix containing the reference signals and $\Omega$ and $\Psi$ being defined as 
\begin{equation}
    \Omega = 
    \begin{bmatrix}
        Q &&& \\
        &Q&& \\
        && \ddots & \\
        &&& P \\
    \end{bmatrix}
    , \Psi = 
    \begin{bmatrix}
        R &&& \\
        &R&& \\
        && \ddots & \\
        &&& R \\
    \end{bmatrix}.  
\end{equation}
This can be cast into quadratic form by
\begin{equation}
    J(x(k),\Delta U_\mathrm{k}) = \frac{1}{2} \Delta U_\mathrm{k}^T G \Delta U_\mathrm{k} + \Delta U_\mathrm{k}^\top F(\Phi x(k) - \Re_\mathrm{k}),
\end{equation}
with 
\begin{equation}
    G = 2(\Psi+\Gamma^\top \Omega \Gamma),
\end{equation}
\begin{equation}
    F = 2 \Gamma^\top \Omega.
\end{equation}
This cost function has an analytical solution for
\begin{equation}
   \Delta U^*_\mathrm{k} = - G^{-1}(F\Phi x(k) - F \Re_\mathrm{k})
\end{equation}
which can be used to obtain the optimal control input of 
\begin{equation}
    u(k) = u(k-1) + \begin{bmatrix}
        1 & 0 & 0 & \dots & 0 \\
    \end{bmatrix}  \Delta U^*_\mathrm{k}.
\end{equation}

\end{document}